\documentclass[12pt,preprint]{aastex}
\begin{document}
\newcommand{\msun}{\mbox{M$_{\odot}$}}
\newcommand{\rsun}{\mbox{R$_{\odot}$}}
\newcommand{\lsun}{\mbox{L$_{\odot}$}}
\title{Variability of Young Massive Stars \\ in the Galactic Super Star
Cluster Westerlund~1}

\author{Alceste Z. Bonanos\altaffilmark{1}}

\altaffiltext{1}{Vera Rubin Fellow, Carnegie Institution of Washington,
Department of Terrestrial Magnetism, 5241 Broad Branch Road NW,
Washington, DC 20015; bonanos@dtm.ciw.edu}

\begin{abstract}

This paper presents the first optical variability study of the
Westerlund~1 super star cluster in search of massive eclipsing binary
systems. A total of 129 new variable stars have been identified,
including the discovery of 4 eclipsing binaries that are cluster
members, 1 additional candidate, 8 field binaries, 19 field $\delta$
Scuti stars, 3 field W~UMa eclipsing binaries, 13 other periodic
variables and 81 long period or non-periodic variables. These include
the known luminous blue variable, the B[e] star, 11 Wolf-Rayet stars,
several supergiants, and other reddened stars that are likely members of
Westerlund~1. The bright X-ray source corresponding to the Wolf-Rayet
star WR77o (B) is found to be a 3.51 day eclipsing binary. The discovery
of a reddened detached eclipsing binary system implies the first
identification of main-sequence stars in Westerlund~1.

\end{abstract}

\keywords{binaries: eclipsing, stars: Wolf-Rayet, stars: variables:
other, open clusters and associations: individual (Westerlund~1)}

\section{Introduction}

Westerlund~1 is a unique laboratory for studying the elusive evolution
of massive stars. It has recently emerged as the nearest known super
star cluster, i.e., a globular cluster precursor, typically found in
starburst galaxies. Its discovery dates back to \citet{Westerlund61},
however, the high amount of reddening ($A_V\sim12$ mag) hindered
spectroscopic observations for many years \citep[see][]{Westerlund87}.
Spectroscopy by \citet{Clark02} and \citet{Clark05} revealed a massive
stellar population of post-main sequence objects and determined it to be
the most massive compact young cluster known in the Local Group.
\citet{Mengel07} measured a dynamical mass of $\log M=4.8$, in agreement
with the mass inferred by \citet{Clark05}.  Westerlund~1 therefore
weighs more than the R136 or Arches clusters, however, with an inferred
age of 4.5-5 Myr \citep{Crowther06}, it is more similar in age to the
Quintuplet and Center cluster. Furthermore, it is less obscured by dust
than the Galactic Center clusters, which allows for observations in the
optical and is ten times nearer than R136, which decreases crowding and
blending effects by the same amount. \citet{Clark05} obtained
spectroscopy of 53 stars, which were all classified as post
main-sequence objects, including blue supergiants, red supergiants,
yellow hypergiants, a luminous blue variable and 14 Wolf-Rayet (WR)
stars. Currently, 24 WR stars have been identified in Westerlund~1
\citep[see][and references therein]{Crowther06}, concentrating $8\%$ of
the known galactic WR stars in one cluster \citep{vanderHucht06}.
$Chandra$ observations have revealed the presence of a magnetar in
Westerlund~1 \citep{Muno06}, which for the first time places a firm
lower limit on the mass of a neutron star progenitor. In addition, 12 WR
stars and a B[e] star were found to be X-ray sources \citep{Skinner06}.

The motivation behind this paper is twofold: to discover eclipsing
binaries containing massive stars ($>50\msun$) and to provide time
domain information for this remarkable cluster. Eclipsing binaries are
extremely powerful tools that can be used to measure the most massive
stars, probe the upper stellar mass limit and provide accurate masses
and radii for the most massive stars in a variety of environments and at
a range of metallicities. Double-lined spectroscopic binary systems
exhibiting eclipses in their light curves provide accurate geometric
measurements of the fundamental parameters of their component
stars. Specifically, the light curve provides the orbital period,
inclination, eccentricity, the fractional radii and flux ratio of the
two stars. The radial velocity semi-amplitudes determine the mass ratio;
the individual masses can be solved using Kepler's third
law. Furthermore, by fitting synthetic spectra to the observed ones, one
can infer the effective temperatures of the stars, solve for their
luminosities and derive the distance \citep[e.g.\@][]{Bonanos06}.

Until recently, the most massive stars measured in eclipsing binaries
were R136-38 in the Large Magellanic Cloud
\citep[$56.9\pm0.6\;\msun$,][]{Massey02} and WR~22
\citep[$55.3\pm7.3\;\msun$,][]{Rauw96, Schweickhardt99}, an evolved star
in our Galaxy. The current heavyweight champion is a galactic WR binary,
WR~20a \citep{Rauw04, Bonanos04}, in the young compact cluster
Westerlund~2, with component masses $83.0 \pm 5.0\; \msun$ and $82.0 \pm
5.0\; \msun$, making this the most massive binary known with an accurate
mass determination. Such systems are of particular interest, since
massive binaries might be progenitors of gamma-ray bursts \citep[GRBs,
e.g.\@][]{Fryer07}, especially in the case of Population III, metal-free
stars \citep[see][]{Bromm06}. They also provide constraints for untested
massive star formation and evolution models, stellar atmosphere and wind
models. Accurate parameters of massive stars provide insights into the
frequency of ``twin'' binaries \citep[see][]{Pinsonneault06,Krumholz06},
the progenitors of X-ray binaries, core-collapse supernovae, the
connection between supernovae and GRBs, and Population III stars, by
studying any observed trends with metallicity.

A systematic search for the most massive stars ($>50\msun$) in eclipsing
binaries is currently underway. Analogs of WR 20a, if not more massive
binaries, are bound to exist in the young massive clusters at the center
of the Galaxy \citep[Center, Arches, Quintuplet; e.g.][]{Peeples07}, in
nearby super star clusters (e.g.\@ Westerlund\,1, NGC 3603 and R136), in
Local Group galaxies (e.g.\@ LMC, SMC, M31, M33) and beyond (e.g.\@ M81,
M83, NGC 2403). The large binary fraction ($>0.6$) measured by
\citet{Kobulnicky06} among early-type stars in the Cygnus OB2
association implies that a significant number of massive short-period
eclipsing binaries should exist in massive clusters. The first step in
measuring accurate masses of the most massive stars involves discovering
eclipsing binaries in massive clusters and nearby galaxies. The
brightest of these are expected to contain very massive stars, and
follow-up spectroscopy provides fundamental parameters for the component
stars.

This paper presents the first variability study of the massive
Westerlund~1 cluster, in search of massive stars in eclipsing binaries.
It is organized as follows: \S 2 describes the observations, \S 3 the
data reduction, calibration and astrometry, \S 4 the variable star
catalog, \S 5 the color magnitude diagram and \S 6 the summary.

\section{Observations}

Photometry of Westerlund~1 was obtained with the Direct CCD on the 1 m
Swope telescope at Las Campanas Observatory, Chile. The camera uses a
$2048\times3150$ SITe CCD with a pixel size of $15\mu \rm m\;
pixel^{-1}$ and a pixel scale of $0\farcs435\; \rm pixel^{-1}$.  A
$1200\times1200$ section of the CCD corresponding to $8.7\arcmin$ on the
side was read out, centered at $\alpha=16\!\!:\!\!47\!\!:\!\!03.8,
\delta=-45\!\!:\!\!50\!\!:\!\!41, {\rm J2000.0}$. The observations were
made on 17 nights between UT 2006 June 15 and July 25, typically in
$IRVB$ sequences. The total number of exposures were 252$\times5\,s$ and
216$\times30\,s$ in the $I$ filter, 217$\times30\,s$ in $R$,
200$\times600\,s$ in $V$, and 167$\times1200\,s$ in $B$. Two exposure
times were taken in $I$ to increase the sensitivity to both bright and
faint cluster members. The median value of the seeing was $1.2\arcsec$
in $I$, $1.3\arcsec$ in $R$, and $1.4\arcsec$ in both $V$ and
$B$. Westerlund~1 was observed at airmasses ranging from 1.04 to 2.69 in
$I$, with the median being 1.16.

\section{Data Reduction, Calibration and Astrometry}

The images were processed with standard IRAF\footnote{IRAF is
distributed by the National Optical Astronomy Observatory, which are
operated by the Association of Universities for Research in Astronomy,
Inc., under cooperative agreement with the NSF.} routines. Specifically,
the images were overscan corrected and trimmed, corrected for the
non-linearity of the CCD with $irred.irlincor$ in IRAF following
\citet{Hamuy06}, and then flat fielded. The value of 23,000 ADU, above
which the non-linearity has not been measured, was adopted to be the
saturation limit in the following reductions.

\subsection{Image Subtraction}
 
The photometry of the variable stars was extracted using the ISIS image
subtraction package \citep{Alard98,Alard00}, which works well in crowded
fields \citep[see][]{Bonanos03}. In each filter, all the frames were
transformed to a common coordinate grid and a reference image was
created by stacking several frames with the best seeing. For each frame,
the reference image was convolved with a kernel to match its PSF and
then subtracted. On the subtracted images, the constant stars cancel
out, and only the signal from variable stars remains. A median image of
all the subtracted images was constructed, and the variable stars were
identified as bright peaks on it by visual inspection. Finally, profile
photometry was extracted from the subtracted images.

DAOPHOT/ALLSTAR PSF photometry \citep{Stetson87} was performed on the
single template images separately and the flux scaling was corrected
when transforming to magnitudes, as described in \citet{Hartman04}. A
more detailed description of the calibration is presented in
\S\ref{photometry}.

\vspace{-0.5cm}
\subsection{Photometric Calibration}
\label{photometry}

On two photometric nights, \citet{Landolt92} standard fields were
observed to calibrate the $BVRI$ light curves to standard
Johnson-Kron-Cousins photometric bands.  Specifically, the SA107, SA
109, PG1323, PG1525 fields were observed on UT 2006 June 19 (N04),
covering a range in airmass from 1.07 to 1.86, and the MarkA, PG0231,
PG1633 fields on UT 2006 July 19 (N12), covering a range in airmass from
1.07 to 2.10.

The transformation from the instrumental to the standard system was
derived using the IRAF $photcal$ package according to the equations:

\vspace{-0.5cm}
\begin{eqnarray*}
b = B + \rm \chi_{b} + \rm \kappa_{b} \cdot X + \rm \xi_{b} \cdot(B-V) \\
v_{1} = V + \rm \chi_{v11} + \rm \kappa_{v12} \cdot X + \rm \xi_{v13} \cdot(B-V) \\
v_{2} = V + \rm \chi_{v21} + \rm \kappa_{v22} \cdot X + \rm \xi_{v23} \cdot(V-I) \\
r = R + \rm \chi_{r} + \rm \kappa_{r} \cdot X + \rm \xi_{r} \cdot(V-R) \\
i = I + \rm \chi_{i} + \rm \kappa_{i} \cdot X + \rm \xi_{i} \cdot(V-I) \\
\end{eqnarray*}
\vspace{-1.5cm}

\noindent where lowercase letters correspond to the instrumental
magnitudes, uppercase letters to standard magnitudes, X is the airmass,
$\chi$ is the zeropoint, $\xi$ the color and $\kappa$ the airmass
coefficient. Additional bright red stars in the observed fields from the
catalog of \citet{Stetson00} increased the color range for N04 to
$-0.22<B-V<1.90$ mag using 95 stars and in N12 to $-0.33<B-V<1.53$ mag
using 58 stars. Table~\ref{calibration} presents both photometric
solutions. The solution of N04 has a larger number of stars, a greater
color range and is in agreement with the values of the $B,V$ color and
extinction coefficients measured by \citet{Hamuy06}; it was therefore
adopted. The transformation was repeated iteratively to determine the
colors of the stars. The lack of observed standard stars with $B-V>2$
mag limits the accuracy of the photometry for the reddest stars. A
comparison with the photometry of P. Stetson (priv. comm.) presented by
\citet{Clark05} shows mean differences (Bonanos$-$Stetson) for the 100
brightest stars in common of $-0.03\pm0.04$ mag in $B$, $-0.01\pm0.06$
mag in $V$, $0.08\pm0.08$ mag in $R$ and $0.24\pm0.10$ mag in $I$. The
large offsets in $R$ and especially in $I$ are due to the lack of very
red standards.  The photometry was therefore scaled by these offsets in
these two bands.

The completeness of the photometry starts to drop rapidly at about 18.5
mag in $I$, 20 mag in $R$, 21.5 mag in $V$ and 22.5 mag in $B$. The CCD
saturates for stars brighter than 10.75 mag in $I$ for the short
exposures and 12.3 mag in the long ones, 12.7 mag in $R$, 15.0 mag in
$V$ and 16.4 mag in $B$. The light curves of variables brighter than
12.3 mag in $I$ were measured from the short exposure images.

The conversion of the variable star light curves to magnitudes involves
two steps. First, it requires a conversion to instrumental magnitudes
that takes into account the aperture correction measured on the template
frame, to achieve correct scaling of fluxes and, consequently, correct
amplitudes for the variables. This requires matching the variables with
stars having DAOPHOT PSF photometry. In the few cases without matches,
the light curves are presented in flux units. The conversion to the
standard system involves an iterative color determination, which in turn
requires detection of the variables in multiple filters. This was not
the case for a large fraction of the stars; therefore, the following
cluster colors $B-V=3.8$ mag, $V-I=5.0$ mag, $V-R=2.8$ mag were adopted
for the conversion, which are average colors for the Westerlund~1 stars
of interest. Hence, the photometry of the variable field stars is
systematically offset by as much as 0.2 mag. Using these colors, the
zeropoints were computed and added to the lightcurves. The $I$ and $R$
light curves were subsequently scaled to Stetson's photometry.

\vspace{-0.9cm}
\subsection{Astrometry}

Equatorial coordinates were determined for the $I$ star list. The
transformation from rectangular to equatorial coordinates was derived
using 940 transformation stars with $V<20$ from the USNO-B1.0
\citep{Monet03} catalog. The median difference between the catalog and
the computed coordinates for over 700 transformation stars was
$0.\!\arcsec2$.

\vspace{-0.5cm}
\section{Variable Star Catalog}

The $BVRI$ light curves were searched separately for variability using
the multiharmonic analysis of variance method \citep{Schwarzenberg96}
and then merged. A total of 129 variables were found in the $8.7\arcmin
\times8.7\arcmin$ region centered on Westerlund~1, many of these being
members of the Westerlund~1 cluster. The variables were classified by
visual inspection into the following five categories: (detached)
eclipsing binaries or (D)EBs for binaries with periods $P>1$ day, W Ursa
Majoris (W~UMa) contact binaries for $P<1$ day and amplitudes $>0.2$
mag, $\delta$ Scuti for short period ($P<0.2$ day) variables with
amplitudes $\sim0.1$ mag, periodic variables ($P>0.2$ day), and
``Other'' that includes the long period or non-periodic
variables. Table~\ref{catalog} presents the variable star catalog.  It
lists the name of the star given by \citet{Westerlund87}, if available,
or the revised identification given by \citet{Clark05}, the RA and Dec
coordinates for J2000.0, the $B$, $V$, $R$, $I$ intensity-averaged
magnitudes, the spectral classification when available from
\citet{Clark05} and for the WR variables from \citet{Crowther06}, the
light curve classification, and period when applicable. Variables with
X-ray detections by \citet{Skinner06} have an ``X'' following their WR
classification. Note, that the catalog includes photometry for some
stars that are missing from \citet{Clark05} due to the gap between their
two CCDs and their smaller field of view. Table~\ref{lcs} presents the
light curves of the variables. The name of each star is based on its
J2000.0 equatorial coordinates, in the format:
$hhmmss.ss$+$ddmmss.s$. In five cases, the light curve is available only
in differential flux units.

The classification results yield 13 eclipsing binaries, 4 of which, and
possibly 5, belong to Westerlund~1, 3 field W~UMa binaries, 19 field
$\delta$ Scuti variables, 13 other variables exhibiting periodic
variability, and 81 variables that are either long period variables or
non-periodic variables. Because of the degeneracy in distinguishing
between W~UMa and $\delta$ Scuti stars from photometry alone, a question
mark is given with these classifications. SX Phe stars are also short
period pulsating stars in the classical instability strip, but belong to
Population~II and are unlikely to be found in the galactic disk.

\begin{figure}[ht]  
\plotone{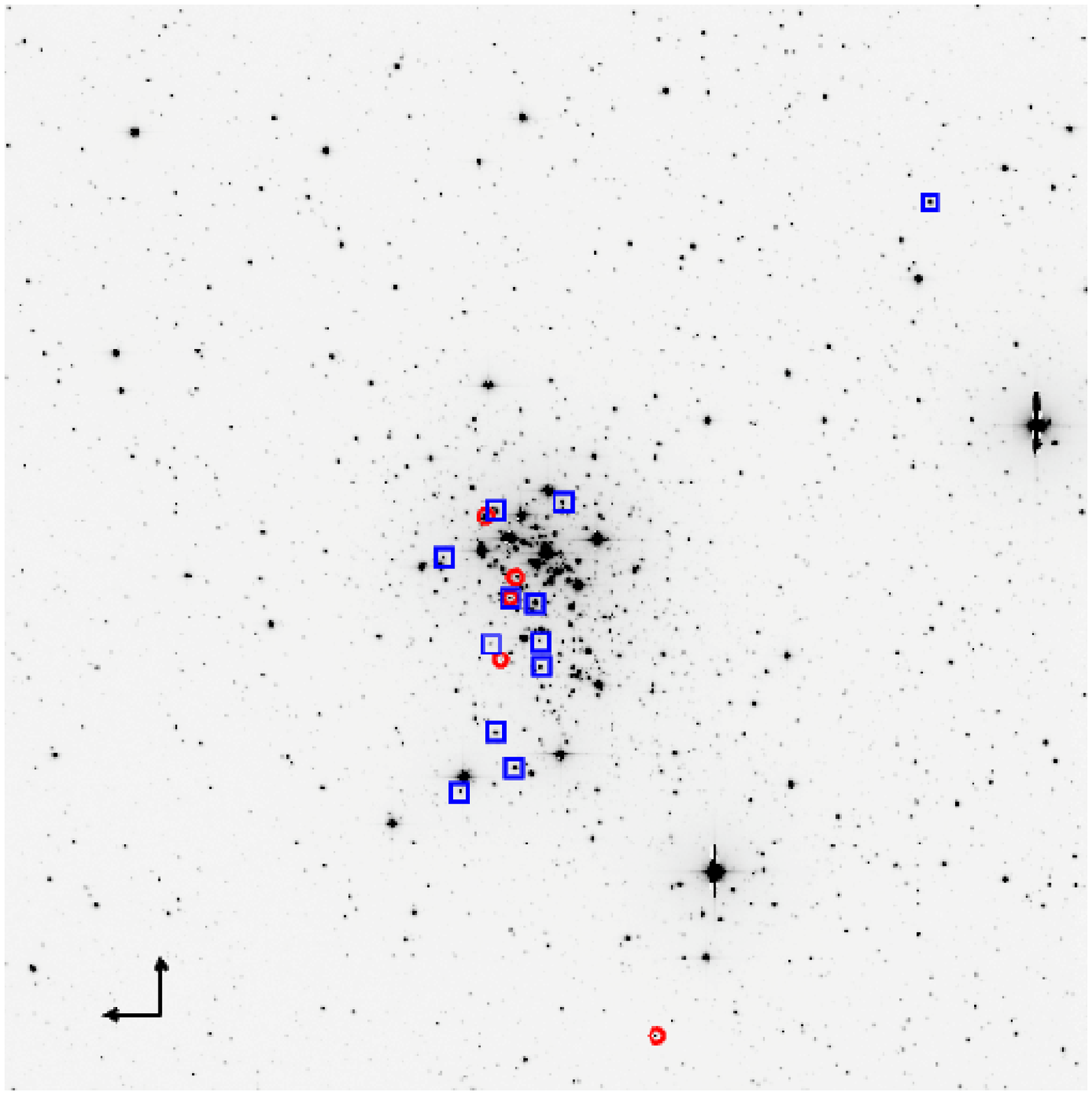}
\caption{$I-$band finderchart of the observed
  $8.7\arcmin\times8.7\arcmin$ field of Westerlund~1 ($0.\!\arcsec9$
  seeing). North is up and East to the left. Cluster eclipsing binaries
  are marked with red circles and variable Wolf-Rayet stars are marked
  with blue squares. Note, the large projected distances of the cluster
  DEB and WR77aa (T).}
\label{finderchart}
\end{figure}

Figure~\ref{finderchart} presents the $I-$band reference image
($0.\!\arcsec9$ seeing) with the positions of the cluster eclipsing
binaries and variable Wolf-Rayet stars. The eclipsing binaries belonging
to Westerlund~1 were selected from the phased light curves as reddened
eclipsing binaries with periods greater than a day. The location of the
reddened DEB and WR77aa (T) demonstrate the extent of the cluster.

\vspace{-0.7cm}
\subsection{Eclipsing Binaries in Westerlund~1}
\vspace{-0.2cm}
\label{ebs}

\begin{figure}[ht]  
\plotone{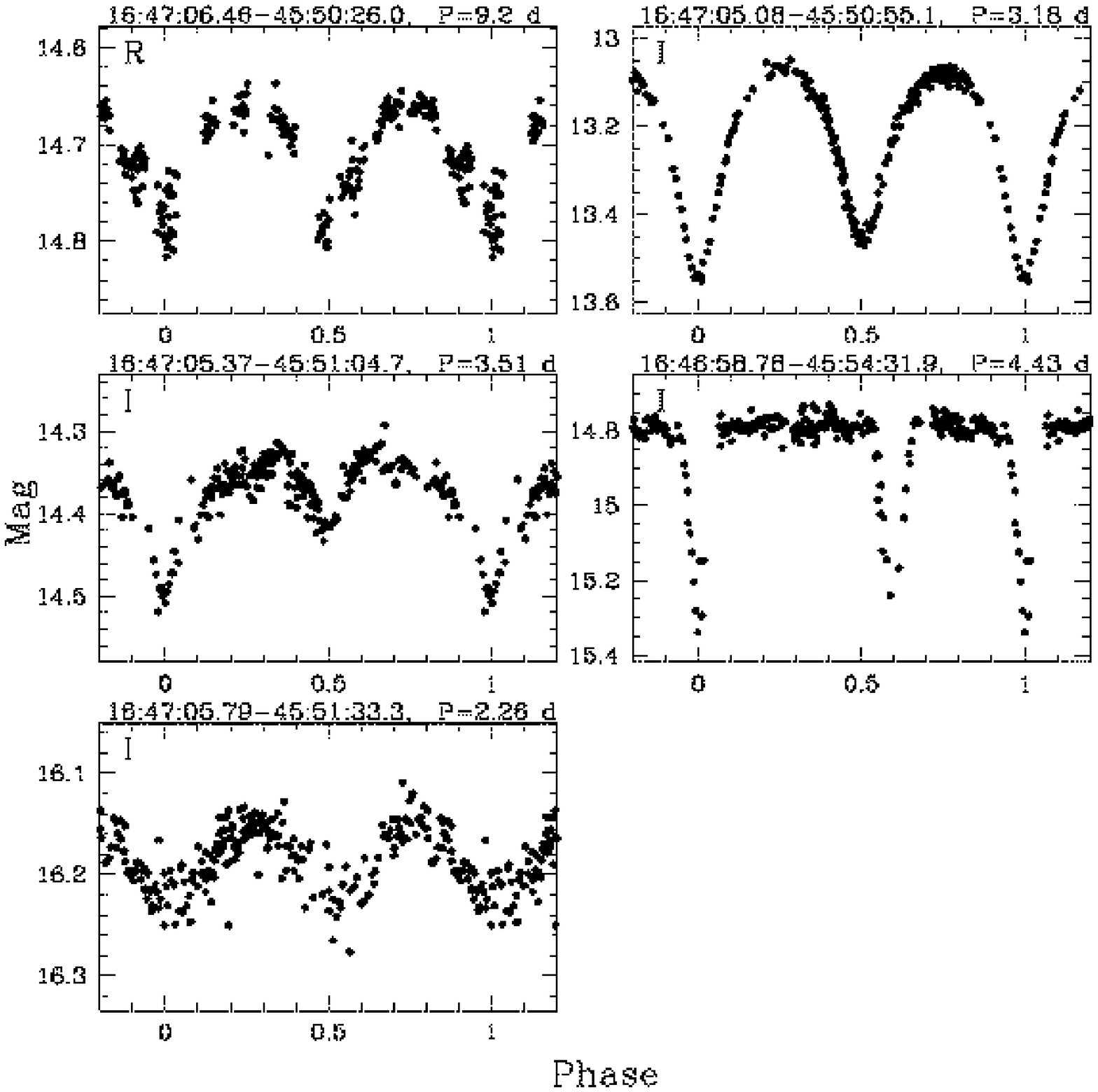}
\caption{Phased light curves of the 5 eclipsing binary members of
Westerlund~1, in order of decreasing brightness in $I-$band. The (RA,
Dec) coordinates, periods and filters are labelled. The 3.51 day system
corresponds to WR77o (B). Cluster membership of the faintest system
needs to be confirmed.}
\label{wdebs}
\end{figure}

Figure~\ref{wdebs} presents phased light curves of the 5 newly
discovered eclipsing binaries in Westerlund~1, in order of decreasing
brightness in $I$. The 9.2 day binary \citep[W13, or star 13 in the
catalog of][]{Westerlund87} is more than a magnitude brighter than the
other eclipsing systems in $I$ (see Table~\ref{catalog}) and thus an
excellent massive star candidate. \citet{Clark05} assigned a
spectroscopic identification of ``OB binary/blend'' for W13, which is in
agreement with the contact eclipsing binary light curve shape. The
second brightest system (W36), with the 3.18 day period, has a light
curve resembling a near-contact configuration that is very accurate,
thus lends itself to accurate parameter determination. There is no
published spectroscopy for W36, but its red color and location, together
with W13, in the core of the cluster give strong evidence for cluster
membership.

The third system, WR77o (B), is a suspected binary from X-ray
observations. \citet{Skinner06} detected strong, hard X-ray emission
with $Chandra$ observations as well as low-amplitude variability in its
X-ray light curve and concluded it is a likely binary. These data
confirm their suspicion, as well as show it to be an eclipsing binary
with a 3.51 day period. The eclipses give it additional value, as they
allow future measurement of the radii and masses of the component stars
of this WN7 system. The uneven out-of-eclipse light curve implies
non-uniform surface brightesses of the component stars, whereas the
uneven eclipse depths imply different temperatures.

The fourth system, a 4.43 day detached eclipsing binary in an eccentric
orbit, is the first main sequence object to be identified in
Westerlund~1, as the detached configuration requires unevolved stars. It
is located far from the cluster core, $\sim4\arcmin$ south of the
cluster, yet its colors are as reddened ($V-I=5.5$ mag) as the cluster
stars, indicating membership in the cluster. Similarly, WR77aa (T), at
an even greater projected distance from the cluster center, is a member
of Westerlund~1 (see Figure~\ref{finderchart}). Measuring the masses of
the above binary component stars will additionally determine the extent
of mass segregation for the most massive stars in this super star
cluster. Finally, the fifth eclipsing binary system found in Westerlund
1 has an extremely reddened color of $R-I=3.7$ mag, is very faint
($R=19.9$ mag), and shows small-amplitude variability with a period of
2.26 days. Cluster membership of this star needs to be confirmed. It is
located near the cluster core, however the $R-$band light curve is too
noisy to check whether the variation in amplitude is the same as in
$I-$band, which is the case for eclipsing systems. It could
alternatively be an extremely reddened pulsating or eclipsing field
star.

It would be premature to draw any conclusions about the binarity
fraction from this data alone, since the monitoring baseline was only
sensitive to short period eclipsing binaries. Spectroscopic monitoring
will reveal additional non-eclipsing systems and, along with X-ray and
radio observations, will yield a more complete estimate of the binary
fraction in Westerlund~1, already estimated by \citet{Crowther06} to be
$>60\%$ for the Wolf-Rayet stars.

\vspace{-1.0cm}
\subsection{Wolf-Rayet Variables in Westerlund~1}
\begin{figure}[ht]  
\plotone{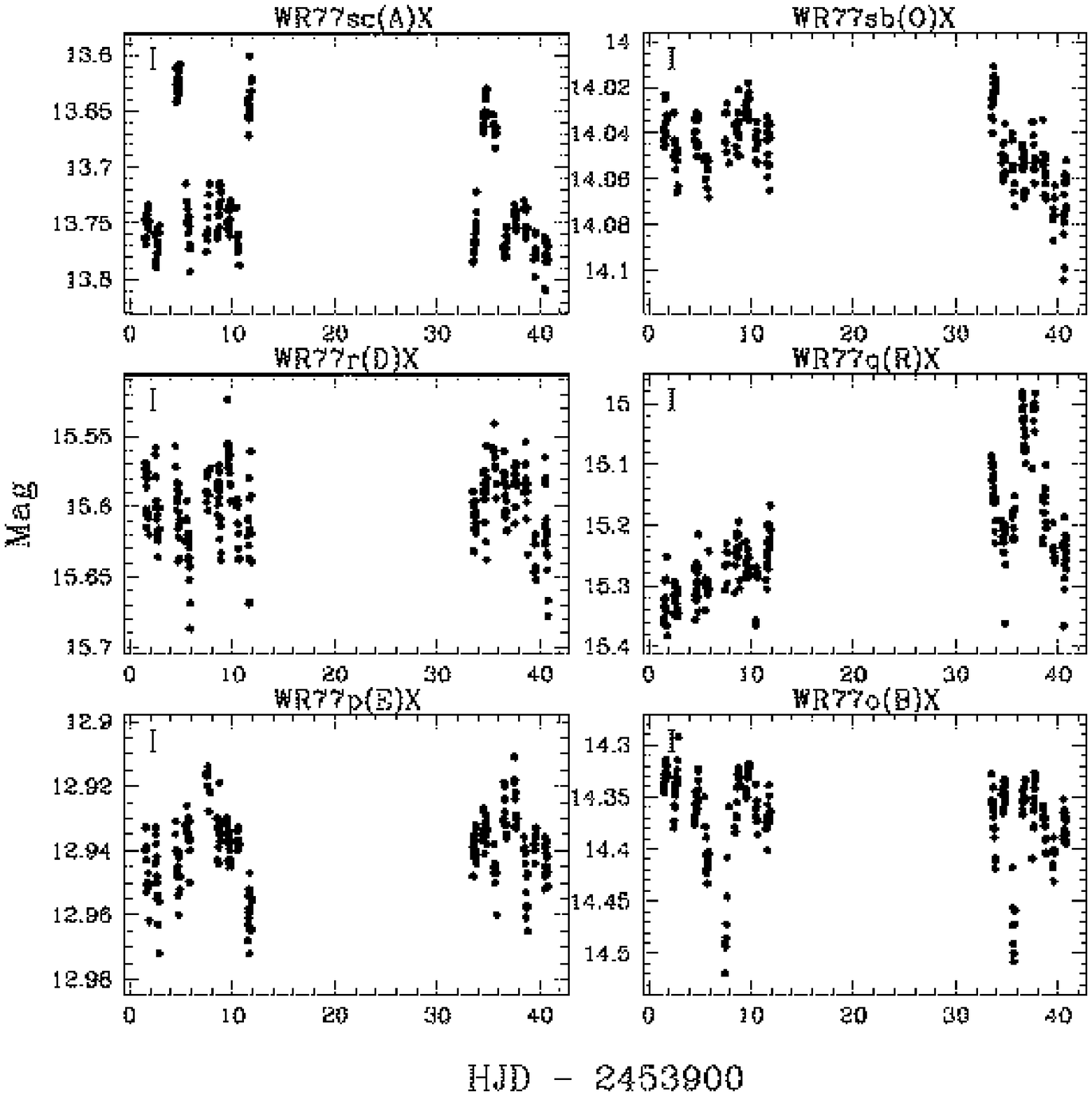}
\caption{Light curves of 6 variable Wolf-Rayet members of
  Westerlund~1. The new name designations \citep{vanderHucht06} are
  followed in parentheses by the letter names \citep{Clark05}. An ``X''
  is appended for sources with X-ray detections from $Chandra$
  \citep{Skinner06}. The plotted filter is labelled in each panel. The
  phased light curve of the eclipsing binary WR77o (B) is shown in
  Figure~\ref{wdebs}. WR77sc (A) roughly phases with a 7.63 day period.}
\label{wr1}
\end{figure}

\begin{figure}[ht]  
\plotone{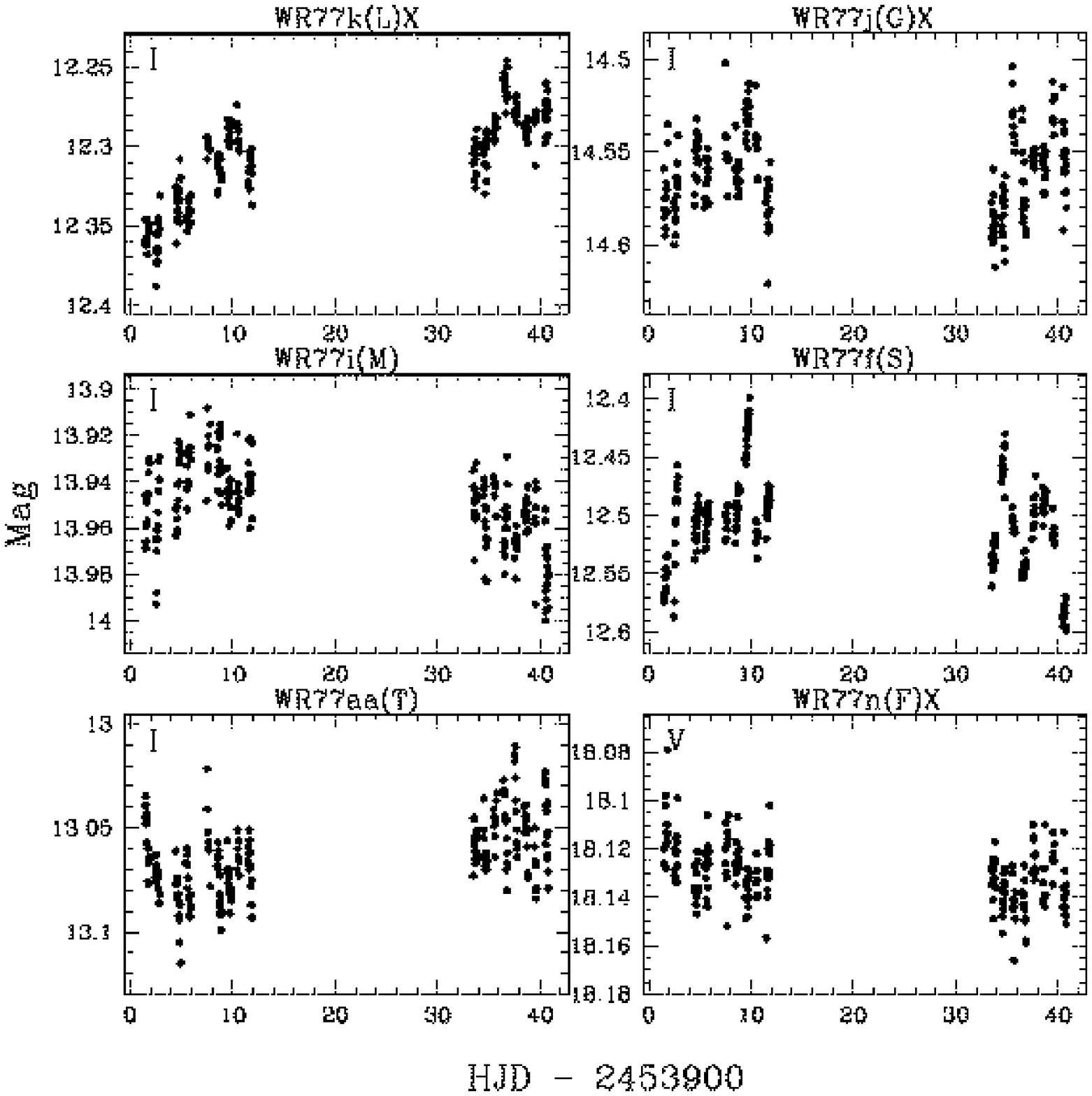}
\caption{Light curves of the other 6 variable Wolf-Rayet members of
  Westerlund~1. Labels are the same as in Figure~\ref{wr1}.}
\label{wr2}
\end{figure}

\vspace{-0.2cm} 
Half of the 24 known Wolf-Rayet variables show significant optical
variability, ranging from 0.05-0.4 mag. Most notably, WR77o (B), a WN7o
type, is a 3.51 day period eclipsing binary with an eclipse depth of 0.2
mag, discussed in \S~\ref{ebs}. The other 11 Wolf-Rayet variable stars
are: WR77sc (A), WR77sb (O), WR77r (D), WR77q (R), WR77p (E), WR77n (F),
WR77k (L), WR77j (G), WR77i (M), WR77f (S), WR77aa
(T). Figures~\ref{wr1} and \ref{wr2} present light curves of the 12
variable WR stars. \citet{Skinner06} have detected X-ray emission in 9
of these 12 variable Wolf-Rayet stars, which correspond to the WR stars
with the largest variability amplitudes and are marked with an ``X'' in
Figures~\ref{wr1} and \ref{wr2}. WR77f (S), with a $\rm
WN10-11h/B0-1Ia+$ spectral type, is an exception. It varies by $\sim$0.2
mag, yet is not an X-ray source. WR77sc (A) shows a sharp 0.15 mag
increase in brightness that roughly phases with a period of 7.63
days. \citet{Skinner06} conclude that both WR77o (B) and WR77sc (A) are
binaries from their high X-ray luminosity. Eclipses are not visible in
the light curve of WR77sc (A), however the periodic variability is
likely related to winds.

The low-amplitude variability (0.05-0.1 mag) observed in the rest of the
WR stars qualitatively resembles the light curve behavior of WR 123
(WN8), monitored by the MOST satellite \citep{Lefevre05} and is likely
due to inhomogeneities in the winds. The present data are not sufficient
for determining whether a periodic signal is present, similar to the 9.8
hour signal found in WR 123. The cause of this periodicity is not well
understood. \citet{Dorfi06} propose that strange-mode pulsations in
hydrogen rich envelopes of stars with high luminosity to mass ratios are
a possible explanation, whereas \citet{Townsend06} argue for a model of
g-mode pulsations driven by a deeper iron opacity bump. Future
monitoring of Westerlund~1, yielding simultanesouly light curves of (at
least) 24 Wolf-Rayet stars, will help in resolving the nature of the
variability.

\vspace{-1.2cm}
\subsection{Other Variables}
\vspace{-0.3cm}

\begin{figure}[ht]  
\plotone{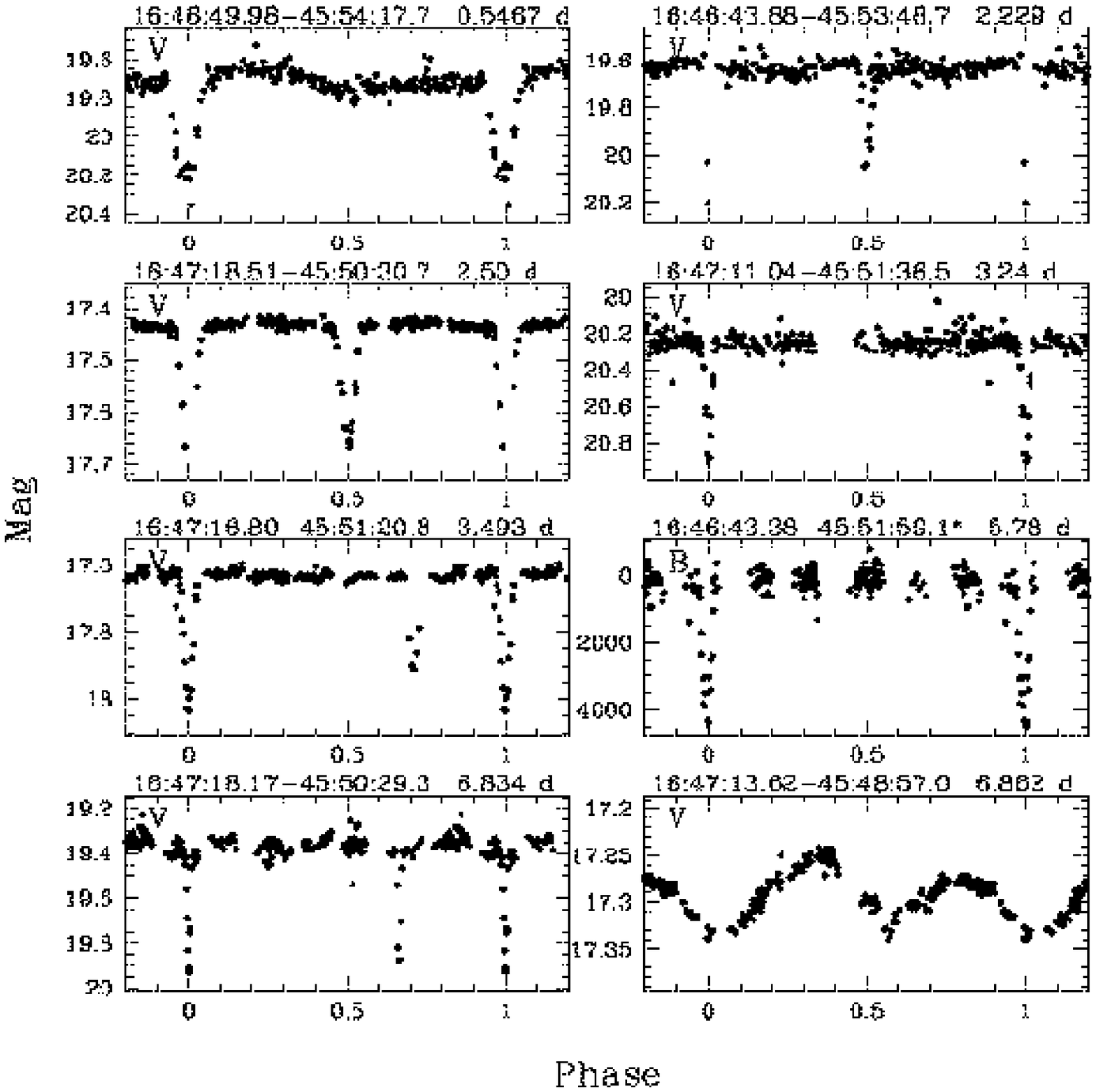}
\caption{Phased light curves of the 7 detached EBs and the RS CVn
binary, in order of increasing periods. Their unreddened colors rule out
cluster membership. The (RA, Dec) coordinates, periods and filters are
labelled. Note that one binary light curve (marked with a *) is given in
flux units from the $B-$band, where it was detected.}
\label{fieldeb}
\end{figure}

\begin{figure}[ht]  
\plotone{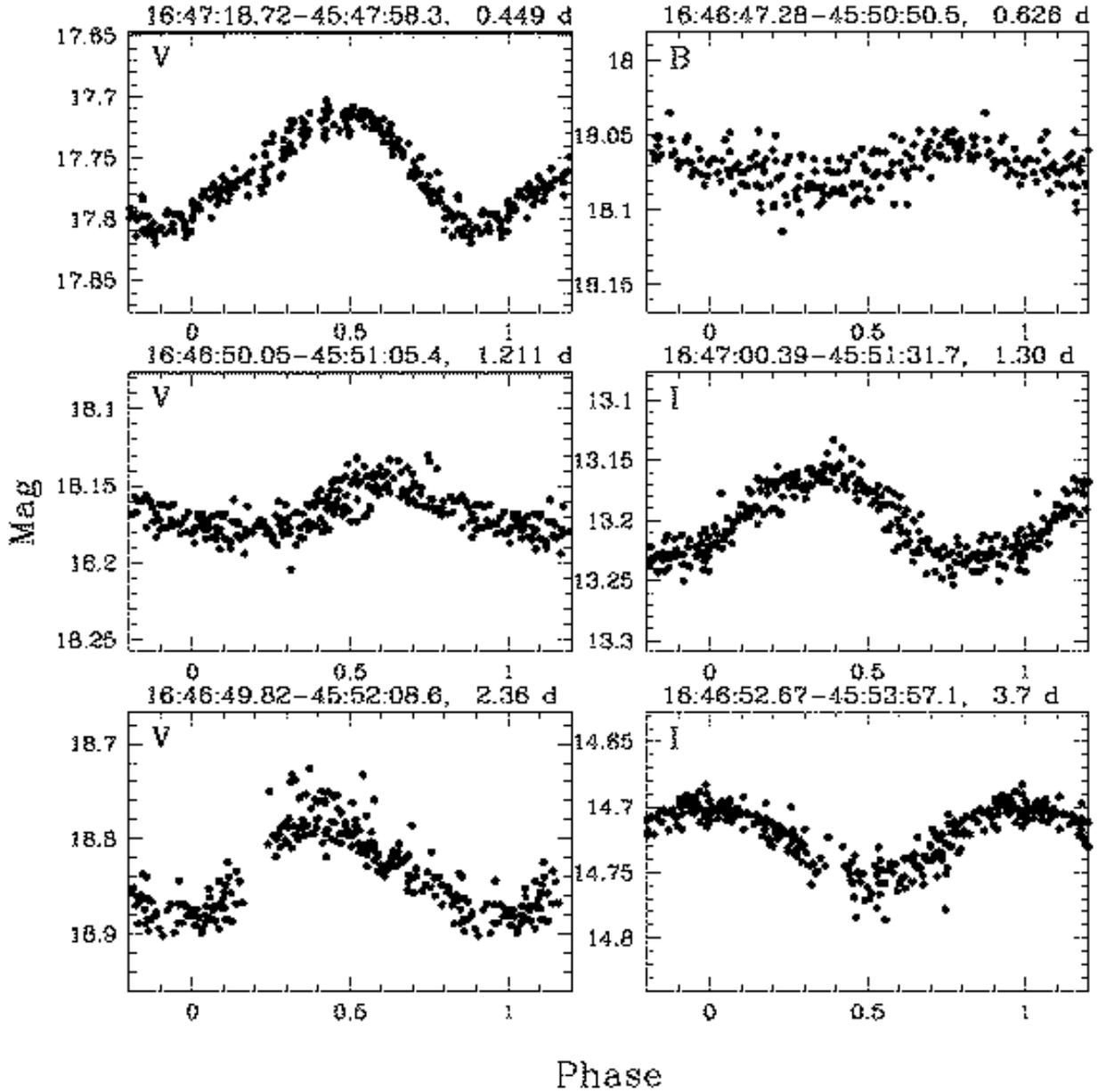}
\caption{Phased light curves of a selection of 6 periodic ($P>0.2$ day)
  small-amplitude variables, possibly due to pulsations. The (RA, Dec)
  coordinates, periods and filters are labelled.}
\label{puls}
\end{figure}

\begin{figure}[ht]  
\plotone{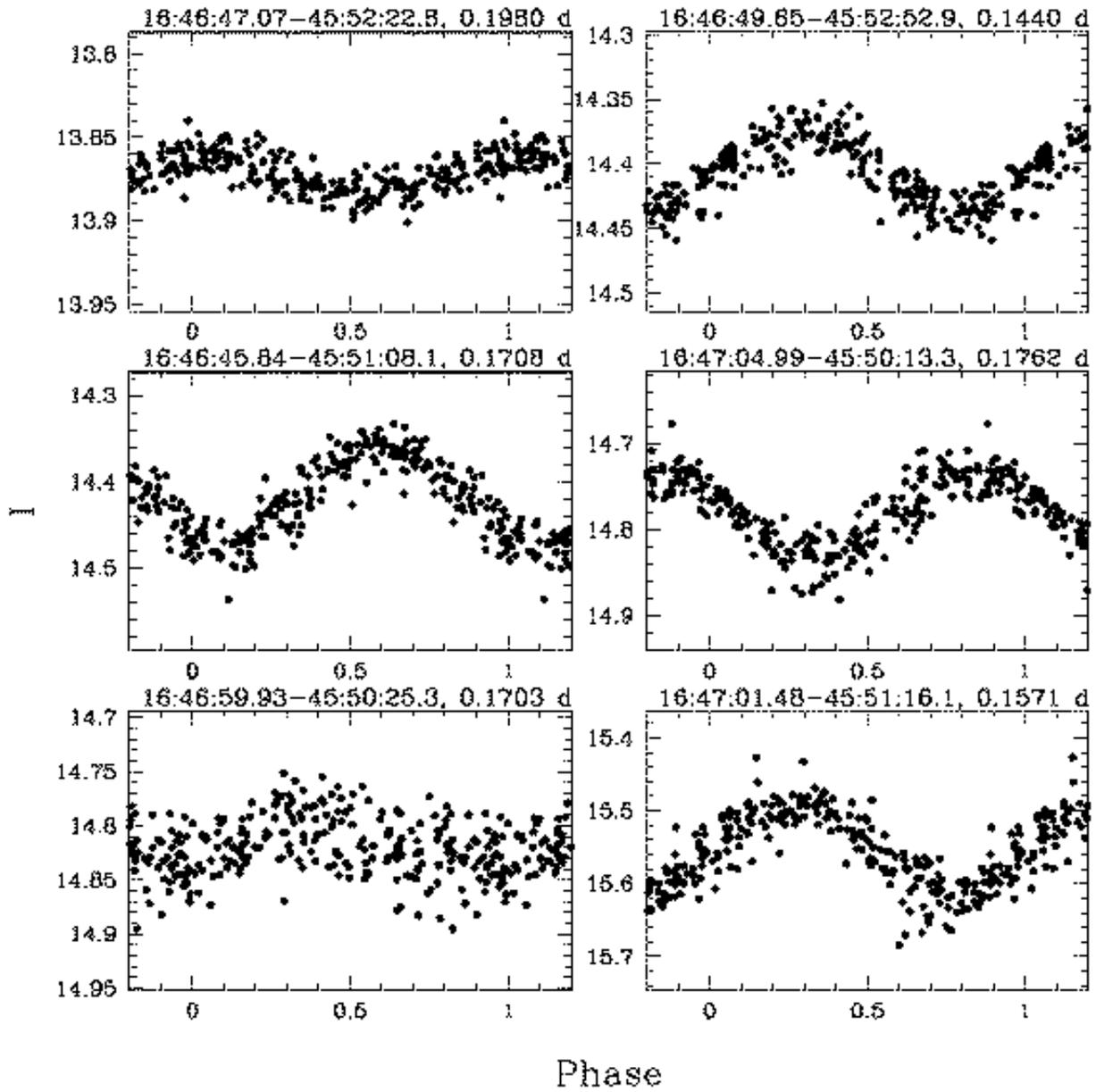}
\caption{Phased $I-$band light curves of a selection of 6 bright field
 $\delta$ Scuti field variables. The (RA, Dec) coordinates and periods
 are labelled. The scatter indicates multiple mode pulsations that are
 typical in such stars.}
\label{dScuti}
\end{figure}

The other periodic variables include 7 field detached eclipsing binary
systems, 2 of which 2 are eccentric, and 1 semi-detached or contact
binary shown in Figure~\ref{fieldeb}. The non-detached field system,
with a 6.862 day period, exhibits an uneven light curve that suggests it
is a magnetically active RS CVn binary. There are also 9 field
small-amplitude variables with $P>0.2$ days (see Figure~\ref{puls}), 19
field small-amplitude variables with $P<0.2$ (see Figure~\ref{dScuti}),
and 4 semi-regular variables. These include WR77sc (A) ($P=7.63$ days)
and W6, an OB supergiant displaying a 2.2 day periodicity with a 0.2 mag
amplitude.  Many of the $\delta$ Scuti variables show significant
scatter in their light curves, indicative of multiple mode pulsations
that are typical in such stars.

\begin{figure}[ht]  
\plotone{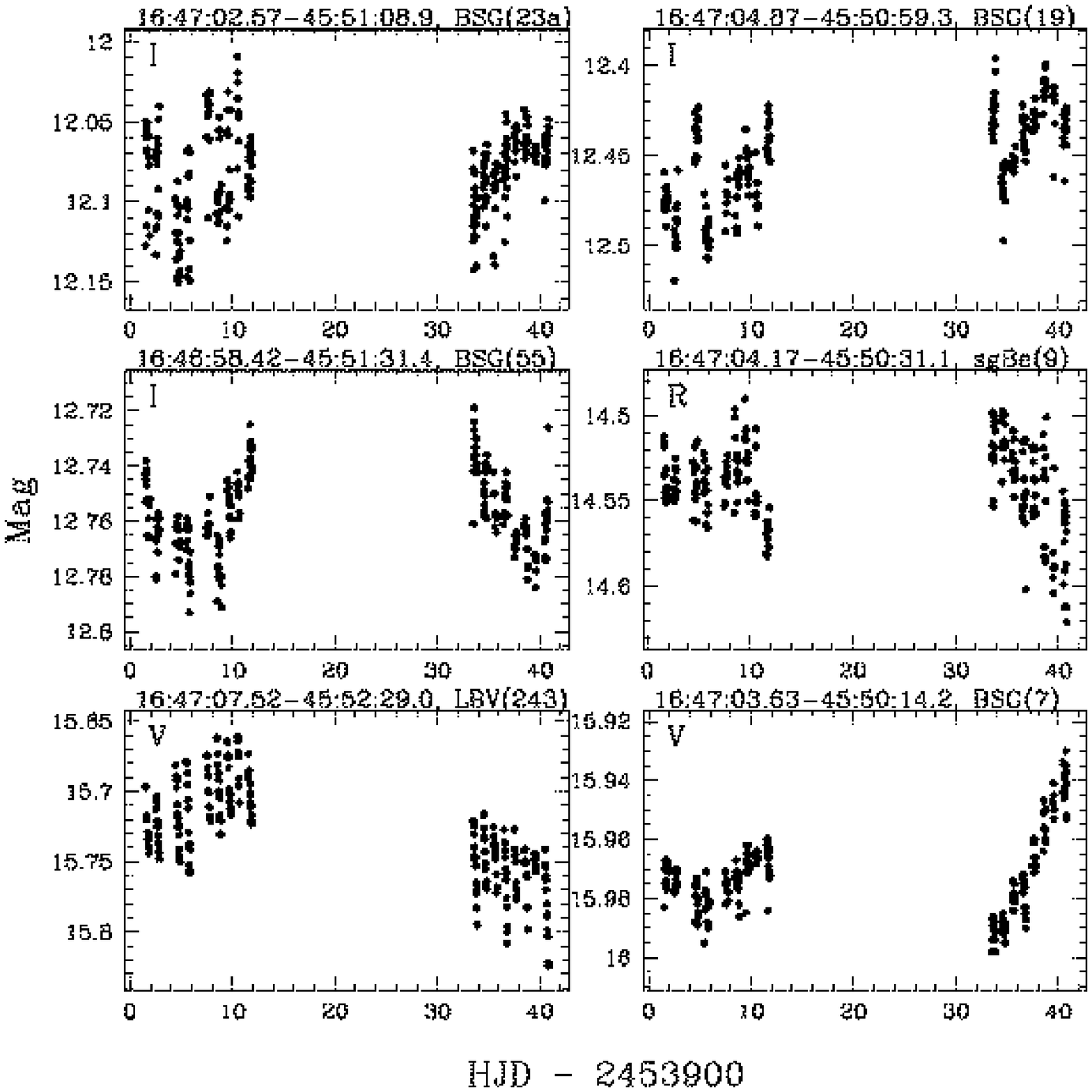}
\caption{Light curves of 6 cluster variables classified as
``Other''. The (RA, Dec) coordinates are followed by the spectral
classification and \citet{Westerlund87} name. The filter is labelled in
the upper left hand corner.}
\label{other}
\end{figure}

A large fraction of the remaining 81 variables classified as ``Other''
have red colors, and can therefore help in identifying additional
supergiants and other massive stars that are cluster
members. Figure~\ref{other} shows light curves of 6 selected cluster
members: the LBV, the B[e] star, and four blue supergiants.

\vspace{-0.8cm}
\section{Color Magnitude Diagram}
\vspace{-0.3cm}

\begin{figure}[ht]  
\plotone{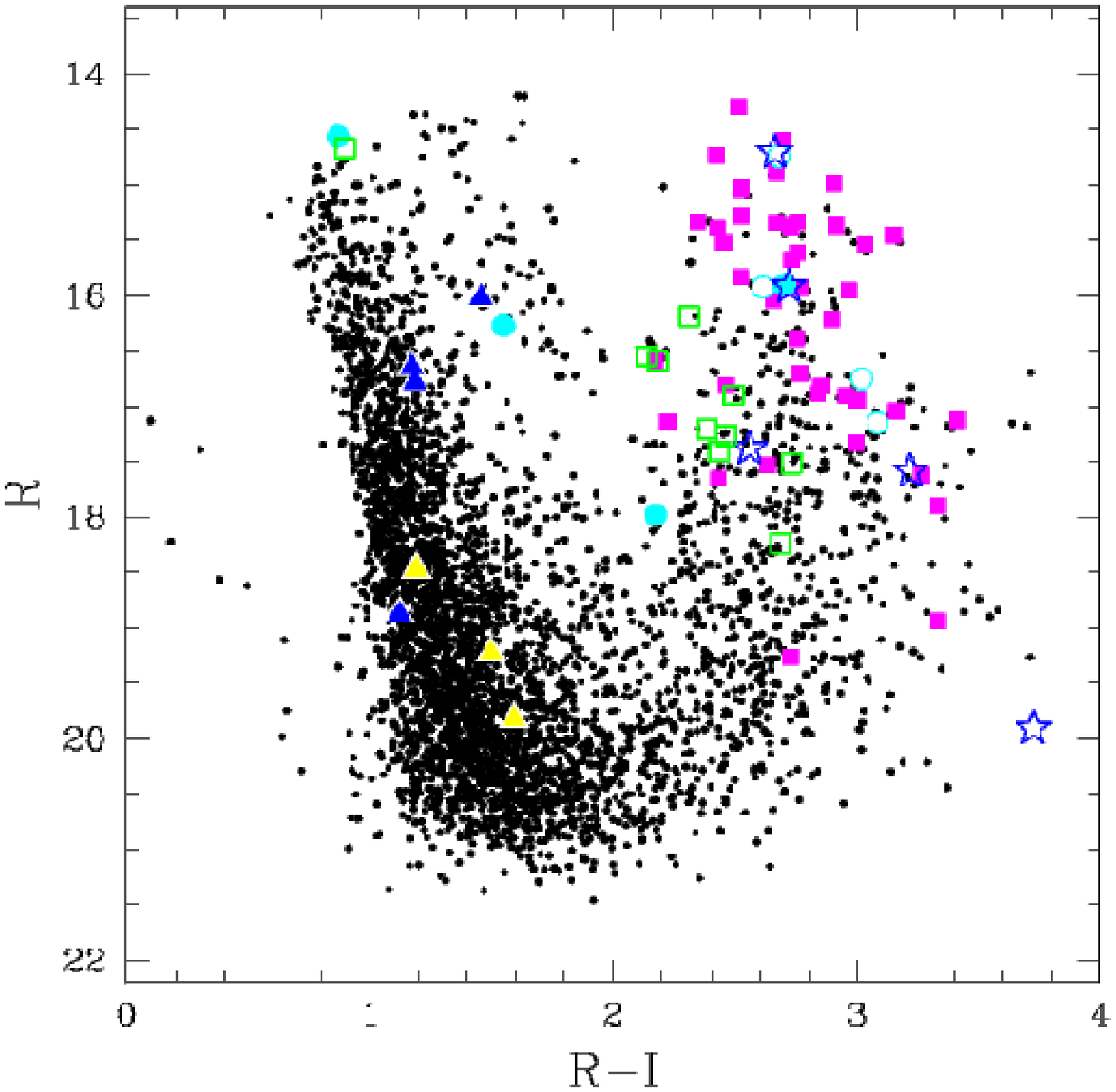}
\caption{$R$ vs.\@ $R-I$ CMD for Westerlund~1 showing the location of
  the variable stars with light curves in these bands. Cluster and field
  eclipsing binaries are marked with blue stars and blue filled
  triangles respectively, $\delta$ Scuti with open green squares, W~UMa
  with yellow filled trianges, other long period and semi-regular
  variables with cyan filled and open circles, respectively. The long
  period or non-periodic ``Other'' variables are marked with magenta
  squares. Many variables, in particular for field stars, were not
  measured in these bands and are therefore missing from the plot.}
\label{rri}
\end{figure}

Figure~\ref{rri} presents the $R$ vs.\@ $R-I$ color-magnitude diagram
(CMD), indicating the positions of the variables and in particular the
eclipsing binaries in the cluster. Field stars are mainly located on the
blue main sequence, whereas the Westerlund 1 stars have reddened colors
between $2.2 < R-I < 3.2$ mag. The sparse second main-sequence
identified by \citet{Clark05} is also observed. It contains several
variables, including the possibly magnetically active RS CVn eclipsing
binary ($P=6.862$ days), which is consistent with this group of stars
being reddened by an intervening absorber. The brightest cluster
eclipsing binary (W13) is remarkably bright, similar in brightness to
the OB supergiants. Although a magnitude fainter, the second brightest
EB (W36) is also located near OB supergiants on the CMD. The cluster DEB
has a similar magnitude to WR77o (B); its position on the CMD is
consistent with it being a cluster member. Spectral classification is
required to confirm membership of the faintest EB. From their position
on the CMD, the 4 semi-regular variables are likely cluster members; 2
of these (W6 and W72) have early-type spectral classifications and the
brightest corresponds to W53.

There is definite contamination from the classical instability strip, as
several $\delta$ Scuti variables occupy the same location as the cluster
members. The 1.3 day variable with a 0.1 mag $I-$band amplitude at
$R-I=2.7, R=15.9$ mag could be a subgiant FK~Comae star. These rapidly
rotating G-K stars are thought to form from coalesced W~UMa binaries
\citep[see][]{Bopp81}. Finally, the bright periodic variables with
periods of 3.7 days ($R-I=1.55, R=16.3$ mag) and 3.715 days ($R-I=0.9,
R=14.6$ mag) could alternatively be low-inclination contact binaries
with 7.4 day periods.
 
\vspace{-0.6cm}
\section{Summary}
\vspace{-0.2cm}

This paper presents the first variability study of the Westerlund~1
cluster and is the first of a series of papers in search of the most
massive stars ($>50\msun$) in eclipsing binaries in young massive
galactic clusters. A total of 129 variable stars are presented,
including cluster supergiant and Wolf-Rayet star light curves, 4 (plus 1
candidate) cluster eclipsing binaries suitable for follow-up
spectroscopy, 8 field eclipsing binaries, 19 pulsating $\delta$ Scuti
variables, 3 W~UMa eclipsing binaries, 13 other periodic variables and
81 long period or non-periodic variables. A significant number of the
long period or non-periodic variables have red colors, suggesting
cluster membership.

The discovery of 4 or 5 eclipsing binaries belonging to Westerlund~1 is
significant, as it provides the means of measuring accurate masses and
radii for these stars. The brightest of these has $V=17.554,\,
R=14.707,\, I=12.044$ mag, therefore its radial velocity curve is
measurable with 4-6 meter class telescopes. Future determination of the
Wolf-Rayet eclipsing binary parameters will be especially useful for
understanding WN7 spectral type stars and probing their evolutionary
state. The cluster detached eclipsing binary is the first main sequence
object to be identified in Westerlund~1 and its location outside the
cluster core may provide insights into cluster formation and evolution.

A systematic search for the most massive stars in eclipsing binaries in
young massive clusters is efficient in discovering massive candidates,
as shown to be the case in Westerlund~1. Radial velocity surveys are
necessary for discovering non-eclipsing systems, however, they require
large amounts of time on 6-8 meter class telescopes to identify binary
candidates and obtain radial velocity curves. Photometric monitoring can
be done using 1-2 meter telescopes and can easily identify short-period
massive eclipsing binary candidates for follow-up spectroscopy.
Eclipsing binary systems are thus extremely valuable probes of the most
massive stars and can yield fundamental parameters of the most massive
stars in both young massive clusters and nearby galaxies.

\acknowledgments{I am very grateful to Craig Heinke for bringing
Westerlund~1 to my attention and thus motivating this variability
study. Many thanks to Peter Stetson for providing his unpublished
photometry of Westerlund~1, to Lucas Macri for help with the photometric
calibration, to Kris Stanek for useful discussions and comments on the
manuscript. I would like to thank Stella Kafka, Phil Massey and Tony
Moffat for a careful reading of the manuscript. I acknowledge research
and travel support from the Carnegie Institution of Washington through a
Vera Rubin Fellowship.}


\clearpage
\begin{deluxetable}{lcccc}
\footnotesize \tablewidth{11.cm} \tablecaption{Transformation
coefficients for the Swope CCD.\label{tab:coeff}} \tablecolumns{4} \tablehead{
\colhead{Instr. Mag.} & \colhead{$\chi$} & \colhead{$\kappa$}
&\colhead{$\xi$}}
\startdata
b$_{N04}$(B-V).....  & $2.981 \pm 0.009$ & $0.205 \pm 0.008$ & $-0.058 \pm 0.005$  \\
v$_{N04}$(B-V).....  & $2.892 \pm 0.009$ & $0.128 \pm 0.007$ & $ 0.056 \pm 0.005$  \\
v$_{N04}$(V-I).....  & $2.888 \pm 0.010$ & $0.130 \pm 0.008$ & $ 0.053 \pm 0.005$  \\
r$_{N04}$(V-R).....  & $2.766 \pm 0.021$ & $0.075 \pm 0.017$ & $-0.051 \pm 0.014$  \\ 
i$_{N04}$(V-I).....  & $3.246 \pm 0.009$ & $0.055 \pm 0.007$ & $-0.042 \pm 0.004$  \\ 
\tableline                                                    
b$_{N12}$(B-V).....  & $3.017 \pm 0.016$ & $0.218 \pm 0.010$ & $-0.041 \pm 0.006$  \\
v$_{N12}$(B-V).....  & $2.890 \pm 0.016$ & $0.161 \pm 0.010$ & $ 0.087 \pm 0.007$  \\
v$_{N12}$(V-I).....  & $2.886 \pm 0.016$ & $0.164 \pm 0.010$ & $ 0.071 \pm 0.006$  \\
r$_{N12}$(V-R).....  & $2.789 \pm 0.016$ & $0.069 \pm 0.010$ & $ 0.007 \pm 0.010$  \\ 
i$_{N12}$(V-I).....  & $3.189 \pm 0.014$ & $0.092 \pm 0.008$ & $-0.040 \pm 0.005$  \\ 
\enddata
\label{calibration}
\end{deluxetable}

\begin{deluxetable}{rccccccrrr}
\rotate
\tabletypesize{\footnotesize}
\tablewidth{0pc}
\tablecaption{\sc Westerlund 1 Variable Star Catalog}
\tablehead{
\colhead{Name} & \colhead{RA} & \colhead{Dec} & \colhead{$\langle B \rangle$} &
\colhead{$\langle V \rangle$} & \colhead{$\langle R \rangle$} &
\colhead{$\langle I \rangle$} & \colhead{Spectr.} & \colhead{Class.}
&\colhead{Period} \\ \colhead{} & \colhead{(2000.0)} &
\colhead{(2000.0)} & \colhead{(mag)} & \colhead{(mag)} & \colhead{(mag)} & \colhead{(mag)} &\colhead{Class..} & \colhead{} &\colhead{(days)}}
\startdata
              ...& 16:47:26.07  &$-$45:50:34.3 &    ...  &  18.794 &    ...  &    ...  &             ... &    Other  &        ... \\
              ...& 16:47:25.67  &$-$45:54:12.9 &  18.881 &  17.333 &    ...  &    ...  &             ... &    Other  &        ... \\
              ...& 16:47:20.83  &$-$45:46:34.1 &    ...  &  15.719 &    ...  &    ...  &             ... &    Other  &        ... \\
              ...& 16:47:20.75  &$-$45:50:02.6 &    ...  &  16.961 &    ...  &    ...  &             ... &    Other  &        ... \\
              ...& 16:47:20.37  &$-$45:48:54.7 &    ...  &  16.523 &    ...  &    ...  &             ... &    Other  &        ... \\
              ...& 16:47:19.50  &$-$45:49:57.3 &    ...  &  17.605 &    ...  &    ...  &             ... &    Other  &        ... \\
              ...& 16:47:18.72  &$-$45:47:58.3 &  19.316 &  17.758 &    ...  &    ...  &             ... &      Per  &   0.449\\
              ...& 16:47:18.51  &$-$45:50:30.7 &  19.227 &  17.450 &  16.646 &  15.474 &             ... &      DEB  &    2.50\\
              ...& 16:47:18.17  &$-$45:50:29.3 &    ...  &  19.387 &    ...  &    ...  &             ... &      DEB  &   6.834\\
              ...& 16:47:16.80  &$-$45:51:20.8 &  19.456 &  17.650 &  16.781 &  15.596 &             ... &      DEB  &   3.493\\
              ...& 16:47:14.19  &$-$45:48:00.7 &    ...  &  18.134 &    ...  &    ...  &             ... &    Other  &        ... \\
              ...& 16:47:13.62  &$-$45:48:57.0 &  19.557 &  17.289 &  16.020 &  14.563 &             ... &       EB  &   6.862\\
              ...& 16:47:13.39  &$-$45:49:10.5 &    ...  &    ...  &  15.960 &  12.990 &             ... &    Other  &        ... \\
              ...& 16:47:12.92  &$-$45:47:54.4 &  21.296 &  19.397 &  18.471 &  17.280 &             ... &    WUMa?  &  0.3657\\
              ...& 16:47:11.60  &$-$45:49:22.4 &    ...  &    ...  &  16.211 &  13.314 &             ... &    Other  &        ... \\
              ...& 16:47:11.06  &$-$45:49:30.9 &    ...  &    ...  &  17.729 &  15.018 &             ... &  $\delta$ Scuti?  &   0.144\\
              ...& 16:47:11.04  &$-$45:51:36.5 &    ...  &  20.268 &    ...  &    ...  &             ... &      DEB  &    3.24\\
              ...& 16:47:09.75  &$-$45:51:24.5 &    ...  &    ...  &  19.702 &  16.694 &             ... &  $\delta$ Scuti?  &   0.134\\
              70 & 16:47:09.39  &$-$45:50:49.6 &    ...  &    ...  &    ...  &  11.352 &           B3Ia  &    Other  & ... \\
              71 & 16:47:08.45  &$-$45:50:49.3 &    ...  &  17.531 &  14.158 &    ...  &           B3Ia  &    Other  &     ... \\
    72,WR77sc(A) & 16:47:08.35  &$-$45:50:45.3 &    ...  &    ...  &  16.750 &  13.728 &         WN7b,X  &      Per  &    7.63\\
              ...& 16:47:08.24  &$-$45:50:56.4 &    ...  &    ...  &  17.123 &  13.710 &             ... &    Other  &        ... \\
              ...& 16:47:07.84  &$-$45:51:47.9 &    ...  &    ...  &  16.940 &  13.939 &             ... &    Other  &        ... \\
       WR77sb(O) & 16:47:07.68  &$-$45:52:35.6 &    ...  &    ...  &  16.886 &  14.047 &         WN6o,X  &    Other  &        ... \\
             243 & 16:47:07.52  &$-$45:52:29.0 &  19.873 &  15.730 &    ...  &    ...  &            LBV  &    Other  &     ... \\
              ...& 16:47:07.29  &$-$45:50:24.2 &    ...  &    ...  &  16.811 &  14.346 &             ... &    Other  &        ... \\
              74 & 16:47:07.08  &$-$45:50:12.9 &    ...  &    ...  &  15.835 &  13.312 &             ... &    Other  &        ... \\
              ...& 16:47:07.01  &$-$45:49:40.1 &    ...  &    ...  &  16.702 &  13.938 &             ... &    Other  &        ... \\
              ...& 16:47:06.68  &$-$45:47:38.5 &    ...  &  18.250 &  14.919 &    ...  &             ... &    Other  &     ... \\
             16a & 16:47:06.63  &$-$45:50:42.1 &    ...  &  16.248 &    ...  &    ...  &          A2Ia+  &    Other  &        ... \\
              15 & 16:47:06.61  &$-$45:50:29.5 &    ...  &  19.301 &    ...  &    ...  &OBbinary/blend?  &    Other  &        ... \\
              13 & 16:47:06.46  &$-$45:50:26.0 &    ...  &  17.554 &  14.707 &  12.044 &OBbinary/blend?  &       EB  &    9.20\\
              ...& 16:47:06.34  &$-$45:48:26.2 &    ...  &  20.486 &  19.230 &  17.733 &             ... &    WUMa?  &   0.371\\
             265 & 16:47:06.29  &$-$45:49:23.7 &    ...  &  17.739 &    ...  &    ...  &          F5Ia+  &    Other  &        ... \\
        WR77r(D) & 16:47:06.28  &$-$45:51:26.4 &    ...  &    ...  &  18.935 &  15.600 &         WN7o,X  &    Other  &        ... \\
              ...& 16:47:06.27  &$-$45:51:03.9 &    ...  &    ...  &  17.150 &  14.065 &             ... &      Per  &     5.2\\
    14c,WR77q(R) & 16:47:06.10  &$-$45:50:22.4 &    ...  &    ...  &  17.652 &  15.223 &         WN5o,X  &    Other  &        ... \\
    241,WR77p(E) & 16:47:06.07  &$-$45:52:08.3 &    ...  &  18.379 &  15.675 &  12.940 &          WC9,X  &    Other  &    ... \\
              25 & 16:47:05.82  &$-$45:50:33.2 &    ...  &  18.205 &  15.331 &  12.575 &             ... &    Other  &     ... \\
              ...& 16:47:05.79  &$-$45:51:33.3 &    ...  &    ...  &  19.907 &  16.182 &             ... &       EB  &    2.26\\
              18 & 16:47:05.73  &$-$45:50:50.4 &    ...  &  17.610 &  14.898 &  12.228 &             ... &    Other  &     ... \\
              69 & 16:47:05.39  &$-$45:51:30.5 &    ...  &  16.322 &    ...  &    ...  &             ... &    Other  &        ... \\
        WR77o(B) & 16:47:05.37  &$-$45:51:04.7 &    ...  &    ...  &  17.588 &  14.367 &         WN7o,X  &       EB  &    3.51\\
    239,WR77n(F) & 16:47:05.21  &$-$45:52:24.8 &    ...  &  18.130 &    ...  &    ...  &         WC9d,X  &    Other  &        ... \\
              ...& 16:47:05.12  &$-$45:51:10.1 &    ...  &    ...  &  17.621 &  14.354 &             ... &    Other  &        ... \\
              36 & 16:47:05.08  &$-$45:50:55.1 &  22.989 &  18.975 &  15.920 &  13.199 &             ... &       EB  &    3.18\\
              ...& 16:47:04.99  &$-$45:50:13.3 &    ...  &    ...  &  17.515 &  14.782 &             ... &  $\delta$ Scuti?  &  0.1762\\
              19 & 16:47:04.87  &$-$45:50:59.3 &    ...  &  18.784 &  15.366 &  12.452 &  O9.5Ia-B0.5Ia  &    Other  &     ... \\
              ...& 16:47:04.77  &$-$45:49:47.1 &  18.568 &  16.896 &    ...  &    ...  &             ... &  $\delta$ Scuti?  &   0.115\\
              8a & 16:47:04.77  &$-$45:50:24.8 &    ...  &  15.913 &    ...  &    ...  &          F5Ia+  &    Other  &        ... \\
              ...& 16:47:04.69  &$-$45:52:06.8 &    ...  &    ...  &  16.899 &  14.405 &             ... &  $\delta$ Scuti?  &  0.1798\\
              28 & 16:47:04.63  &$-$45:50:38.4 &    ...  &  17.146 &  14.303 &    ...  &  O9.5Ia-B0.5Ia  &    Other  &     ... \\
              20 & 16:47:04.68  &$-$45:51:23.9 &    ...  &    ...  &  15.660 &    ...  &          $<$M6I  &   Other    &        ... \\
              ...& 16:47:04.55  &$-$45:50:08.5 &    ...  &    ...  &    ...  &    ...  &             ... &    Other  &    ... \\
              ...& 16:47:04.52  &$-$45:51:19.4 &    ...  &  16.556 &    ...  &    ...  &             ... &    Other  &        ... \\
             238 & 16:47:04.41  &$-$45:52:27.6 &    ...  &  17.820 &    ...  &    ...  &  O9.5Ia-B0.5Ia  &    Other  &        ... \\
     44,WR77k(L) & 16:47:04.21  &$-$45:51:07.1 &    ...  &    ...  &  15.457 &  12.307 &        WN9h:,X  &    Other  &        ... \\
               9 & 16:47:04.17  &$-$45:50:31.1 &    ...  &  18.056 &  14.541 &    ...  &         sgB[e]  &    Other  &     ... \\
              33 & 16:47:04.13  &$-$45:50:48.4 &  20.090 &    ...  &  12.803 &    ...  &          B5Ia+  &    Other  &     ... \\
              60 & 16:47:04.13  &$-$45:51:52.1 &    ...  &    ...  &  16.041 &  13.384 &  O9.5Ia-B0.5Ia  &    Other  &        ... \\
        WR77j(G) & 16:47:04.03  &$-$45:51:25.1 &    ...  &    ...  &  17.893 &  14.561 &        (WN7o)X  &    Other  &        ... \\
     66,WR77i(M) & 16:47:03.96  &$-$45:51:37.6 &    ...  &    ...  &  16.906 &  13.949 &           WC9d  &    Other  &        ... \\
              ...& 16:47:03.96  &$-$45:52:12.0 &    ...  &    ...  &  17.262 &  14.797 &             ... &  $\delta$ Scuti?  &  0.1337\\
              46 & 16:47:03.93  &$-$45:51:19.6 &    ...  &    ...  &  15.537 &  12.499 &             ... &    Other  &        ... \\
              ...& 16:47:03.75  &$-$45:51:12.7 &    ...  &    ...  &  17.042 &  13.879 &             ... &    Other  &        ... \\
               7 & 16:47:03.63  &$-$45:50:14.2 &    ...  &  15.972 &    ...  &    ...  &          B5Ia+  &    Other  &        ... \\
             43a & 16:47:03.56  &$-$45:50:57.6 &    ...  &    ...  &  15.378 &  12.647 &  O9.5Ia-B0.5Ia  &    Other  &        ... \\
              10 & 16:47:03.35  &$-$45:50:34.6 &  22.720 &  18.516 &  15.257 &    ...  &  O9.5Ia-B0.5Ia  &    Other  & ... \\
             42a & 16:47:03.25  &$-$45:50:52.1 &    ...  &    ...  &    ...  &  10.764 &         B5Ia+?  &    Other  & ... \\
              ...& 16:47:03.24  &$-$45:53:11.1 &    ...  &    ...  &  18.545 &  15.847 &             ... &  $\delta$ Scuti?  &  0.1581\\
              ...& 16:47:03.11  &$-$45:51:31.1 &    ...  &  19.167 &  15.929 &  13.165 &             ... &    Other  &     ... \\
             237 & 16:47:03.10  &$-$45:52:18.8 &    ...  &  19.008 &  13.634 &    ...  &           <M6I  &    Other  &     ... \\
               6 & 16:47:03.06  &$-$45:50:23.8 &    ...  &  18.814 &  15.928 &  13.313 &  O9.5Ia-B0.5Ia  &      Per  &    2.20\\
      5,WR77f(S) & 16:47:02.96  &$-$45:50:19.7 &  21.552 &  17.792 &  15.035 &  12.508 &WN10-11h/B0-1Ia+ &     Other &  ... \\
              ...& 16:47:02.71  &$-$45:50:23.9 &  19.651 &    ...  &    ...  &    ...  &             ... &    Other  &        ... \\
              ...& 16:47:02.69  &$-$45:50:50.3 &    ...  &    ...  &  15.340 &  12.669 &             ... &    Other  &        ... \\
             23a & 16:47:02.57  &$-$45:51:08.9 &    ...  &  18.316 &  14.986 &  12.081 &  O9.5Ia-B0.5Ia  &    Other  &   ... \\
              48 & 16:47:02.46  &$-$45:51:24.9 &    ...  &    ...  &  16.818 &  13.966 &             ... &    Other  &        ... \\
              ...& 16:47:02.32  &$-$45:51:15.2 &    ...  &    ...  &  17.536 &  14.902 &             ... &    Other  &        ... \\
             61a & 16:47:02.30  &$-$45:51:41.8 &    ...  &  17.506 &  14.633 &    ...  &  O9.5Ia-B0.5Ia  &    Other  &     ... \\
              11 & 16:47:02.25  &$-$45:50:47.1 &    ...  &  17.508 &  14.599 &  11.898 &  O9.5Ia-B0.5Ia  &    Other  & \\
             12a & 16:47:02.23  &$-$45:50:59.0 &    ...  &  17.575 &  13.611 &    ...  &          A5Ia+  &    Other  &     ... \\
              49 & 16:47:01.91  &$-$45:50:31.6 &    ...  &    ...  &  16.392 &  13.638 &             ... &    Other  &        ... \\
              52 & 16:47:01.85  &$-$45:51:29.4 &    ...  &  17.876 &  14.745 &  12.065 &             ... &      Per  &     6.7\\
              ...& 16:47:01.69  &$-$45:52:57.8 &    ...  &    ...  &  15.393 &  12.966 &             ... &    Other  &        ... \\
              78 & 16:47:01.56  &$-$45:49:57.9 &    ...  &  17.413 &  14.621 &    ...  &             ... &    Other  &     \\
              ...& 16:47:01.48  &$-$45:51:16.1 &    ...  &    ...  &  18.240 &  15.553 &             ... &  $\delta$ Scuti?  &  0.1571\\
               4 & 16:47:01.42  &$-$45:50:37.4 &  18.781 &    ...  &    ...  &    ...  &          F2Ia+  &    Other  &        ... \\
              ...& 16:47:01.44  &$-$45:52:35.0 &    ...  &    ...  &  15.334 &  12.989 &             ... &    Other  &        ... \\
             57a & 16:47:01.36  &$-$45:51:45.6 &    ...  &  16.877 &    ...  &    ...  &           B3Ia  &    Other  &  \\
              21 & 16:47:01.13  &$-$45:51:13.4 &    ...  &    ...  &  15.619 &  12.861 &             ... &    Other  &        ... \\
              ...& 16:47:00.88  &$-$45:51:20.4 &    ...  &    ...  &  17.336 &  14.340 &             ... &    Other  &        ... \\
              ...& 16:47:00.52  &$-$45:48:29.7 &    ...  &    ...  &  17.981 &  15.806 &             ... &      Per  &  1.089\\
              ...& 16:47:00.54  &$-$45:52:23.3 &  17.905 &    ...  &    ...  &    ...  &             ... &    Other  &        ... \\
              53 & 16:47:00.39  &$-$45:51:31.7 &    ...  &  18.950 &  15.901 &  13.198 &             ... &      Per  &    1.30\\
              ...& 16:46:59.93  &$-$45:50:25.3 &    ...  &    ...  &  17.206 &  14.818 &             ... &  $\delta$ Scuti?  &  0.1703\\
              ...& 16:46:59.77  &$-$45:51:40.0 &    ...  &    ...  &  19.264 &  16.536 &             ... &    Other  &        ... \\
              2a & 16:46:59.71  &$-$45:50:51.2 &    ...  &  16.978 &  14.287 &  11.771 &  O9.5Ia-B0.5Ia  &    Other  & ... \\
              ...& 16:46:58.97  &$-$45:50:17.8 &  19.188 &  17.650 &    ...  &    ...  &             ... &    Other  &     ... \\
              ...& 16:46:58.78  &$-$45:54:31.9 &    ...  &  20.313 &  17.377 &  14.816 &             ... &      DEB  &    4.43\\
              55 & 16:46:58.42  &$-$45:51:31.4 &    ...  &  18.030 &  15.283 &  12.758 &  O9.5Ia-B0.5Ia  &    Other  &     ... \\
              ...& 16:46:57.71  &$-$45:53:20.0 &    ...  &    ...  &  14.741 &  12.318 &             ... &    Other  &   \\
              ...& 16:46:56.89  &$-$45:52:04.4 &    ...  &    ...  &  17.141 &  14.917 &             ... &    Other  &        ... \\
              ...& 16:46:54.45  &$-$45:53:30.0 &    ...  &    ...  &  17.408 &  14.975 &             ... &  $\delta$ Scuti?  &  0.1271\\
              ...& 16:46:53.44  &$-$45:53:00.3 &    ...  &  18.061 &  16.598 &  14.425 &             ... &    Other  &     \\
              ...& 16:46:52.67  &$-$45:53:57.1 &    ...  &  16.804 &  16.274 &  14.724 &             ... &      Per  &     3.7\\
              ...& 16:46:52.24  &$-$45:51:15.5 &  17.941 &  16.546 &    ...  &    ...  &             ... &    Other  &     \\
              ...& 16:46:51.70  &$-$45:49:07.9 &    ...  &  16.289 &    ...  &    ...  &             ... &  $\delta$ Scuti?  &  0.4852\\
              ...& 16:46:50.83  &$-$45:53:02.7 &    ...  &    ...  &  14.568 &  13.700 &             ... &      Per  &   3.715\\
              ...& 16:46:50.05  &$-$45:51:05.4 &    ...  &  18.165 &    ...  &    ...  &             ... &      Per  &   1.211\\
              ...& 16:46:49.98  &$-$45:54:17.7 &  21.469 &  19.730 &  18.874 &  17.752 &             ... &       EB  &  0.5467\\
              ...& 16:46:49.82  &$-$45:52:08.6 &    ...  &  18.825 &    ...  &    ...  &             ... &      Per  &    2.36\\
              ...& 16:46:49.65  &$-$45:52:52.9 &    ...  &  18.887 &  16.589 &  14.406 &             ... &  $\delta$ Scuti?  &  0.1440\\
              ...& 16:46:48.38  &$-$45:53:56.6 &    ...  &    ...  &  18.324 &  15.682 &             ... &  $\delta$ Scuti?  &  0.1381\\
              ...& 16:46:47.28  &$-$45:50:50.5 &  18.071 &    ...  &    ...  &    ...  &             ... &      Per  &   0.626\\
              ...& 16:46:47.07  &$-$45:52:22.8 &    ...  &    ...  &  16.187 &  13.872 &             ... &  $\delta$ Scuti?  &  0.1980\\
       WR77aa(T) & 16:46:46.30  &$-$45:47:58.2 &    ...  &    ...  &  15.516 &  13.063 &           WC9d  &    Other  &        ... \\
              ...& 16:46:46.08  &$-$45:48:44.3 &    ...  &  17.372 &    ...  &    ...  &             ... &    Other  &        ... \\
              ...& 16:46:45.84  &$-$45:51:08.1 &    ...  &  18.808 &  16.548 &  14.415 &             ... &  $\delta$ Scuti?  &  0.1708\\
              ...& 16:46:43.88  &$-$45:53:48.7 &    ...  &  19.646 &    ...  &    ...  &             ... &      DEB  &   2.229\\
              ...& 16:46:43.38  &$-$45:51:59.1 &    ...  &    ...  &    ...  &    ...  &             ... &      DEB  &    5.78\\
              ...& 16:46:42.88  &$-$45:50:39.9 &  18.383 &    ...  &    ...  &    ...  &             ... &    Other  &        ... \\
              ...& 16:46:42.32  &$-$45:51:50.4 &    ...  &  18.115 &    ...  &    ...  &             ... &    Other  &        ... \\
              ...& 16:46:42.10  &$-$45:47:01.7 &    ...  &  21.391 &  19.826 &  18.233 &             ... &    WUMa?  &   0.398\\
              ...& 16:46:41.74  &$-$45:47:13.8 &    ...  &  16.647 &  14.673 &  13.773 &             ... &  $\delta$ Scuti?  &  0.1642\\
              ...& 16:46:41.65  &$-$45:46:41.1 &  17.994 &    ...  &    ...  &    ...  &             ... &  $\delta$ Scuti?  &  0.1841\\
              ...& 16:46:40.03  &$-$45:47:50.4 &    ...  &    ...  &    ...  &    ...  &             ... &  $\delta$ Scuti?  &  0.1650\\
              ...& 16:46:40.08  &$-$45:51:07.8 &    ...  &  18.285 &    ...  &    ...  &             ... &      Per  &   0.738\\
              ...& 16:46:39.82  &$-$45:47:46.4 &  16.940 &    ...  &    ...  &    ...  &             ... &  $\delta$ Scuti?  &  0.1642\\
\enddata                         
\label{catalog}
\end{deluxetable} 

\begin{deluxetable}{ccccc}
\tabletypesize{\footnotesize}
\tablewidth{0pc}
\tablecaption{\sc Light Curves of Variables in Westerlund~1}
\tablehead{
\colhead{Name} & \colhead{Filter} & \colhead{HJD$-$2450000} &
\colhead{mag/flux} & \colhead{$\sigma_{mag}/\sigma_{flux}$}}
\startdata
16:47:26.07$-$45:50:34.3 & V & 3901.50711 &     18.798 &   0.013 \\ 
16:47:26.07$-$45:50:34.3 & V & 3901.53865 &     18.776 &   0.011 \\
16:47:26.07$-$45:50:34.3 & V & 3901.59942 &     18.808 &   0.018 \\ 
16:47:26.07$-$45:50:34.3 & V & 3901.62506 &     18.779 &   0.018 \\ 
16:47:26.07$-$45:50:34.3 & V & 3901.65933 &     18.778 &   0.019 \\ 
16:47:26.07$-$45:50:34.3 & V & 3901.70142 &     18.776 &   0.019 \\ 
16:47:26.07$-$45:50:34.3 & V & 3901.74311 &     18.788 &   0.020 \\ 
16:47:26.07$-$45:50:34.3 & V & 3901.77163 &     18.770 &   0.019 \\ 
16:47:26.07$-$45:50:34.3 & V & 3901.79786 &     18.774 &   0.023 \\ 
\enddata                         
\tablecomments{Table 3 is available in its entirety in the electronic
version of the Astronomical Journal. A portion is shown here for
guidance regarding its form and content.}
\label{lcs}
\end{deluxetable} 

\end{document}